\documentclass[twocolumn,aps,pra,showpacs,groupedaddress,superscriptaddress,nofootinbib]{revtex4-1}

\usepackage[pdftex,linkcolor=blue,citecolor=blue,urlcolor=blue,colorlinks]{hyperref}
\usepackage[dvipsnames]{xcolor}
\usepackage{graphicx,epsfig}
\usepackage{braket}
\usepackage{color}
\usepackage{cancel}
\usepackage{inputenc}
\usepackage{amsmath}
\usepackage{amsfonts}
\usepackage{fancyhdr}

\begin{document}
\title{Topological edge states with ultracold atoms carrying orbital angular momentum in a diamond chain}
\author{G. Pelegr\'i}
\affiliation{Departament de F\'isica, Universitat Aut\`onoma de Barcelona, E-08193 Bellaterra, Spain.}
\author{A. M. Marques}
\affiliation{Department of Physics and I3N, University of Aveiro, 3810-193 Aveiro, Portugal.}
\author{R. G. Dias}
\affiliation{Department of Physics and I3N, University of Aveiro, 3810-193 Aveiro, Portugal.}
\author{A. J. Daley}
\affiliation{Department of Physics and SUPA, University of Strathclyde, Glasgow G4 0NG, United Kingdom.}
\author{V. Ahufinger}
\affiliation{Departament de F\'isica, Universitat Aut\`onoma de Barcelona, E-08193 Bellaterra, Spain.}
\author{J. Mompart}
\affiliation{Departament de F\'isica, Universitat Aut\`onoma de Barcelona, E-08193 Bellaterra, Spain.}

\begin{abstract}
We study the single-particle properties of a system formed by ultracold atoms loaded into the manifold of $l=1$ Orbital Angular Momentum (OAM) states of an optical lattice with a diamond chain geometry. Through a series of successive basis rotations, we show that the OAM degree of freedom induces phases in some tunneling amplitudes of the tight-binding model that are equivalent to a net $\pi$ flux through the plaquettes and give rise to a topologically non-trivial band structure and protected edge states. In addition, we demonstrate that quantum interferences between the different tunneling processes involved in the dynamics may lead to Aharanov-Bohm caging in the system. All these analytical results are confirmed by exact diagonalization numerical calculations.
\end{abstract}
\pacs{}
\maketitle


\section{Introduction}
Since the observation of the quantum Hall effect in two-dimensional electron gases \cite{QuantumHallInt,QuantumHallFrac} and the discovery of its relation with topology \cite{QuantumHallTop}, the study of systems with non-trivial topological properties has become a central topic in condensed matter physics. A very interesting example of such exotic phases of matter are topological insulators \cite{TopInsReview}, which are materials that exhibit insulating properties on their bulk but have topologically protected conducting states on their edges. There are many different types of topological insulators, which can be systematically classified in terms of their symmetries and dimensionality \cite{TopInsSymReview}. 
\\
\\
In recent years, many efforts have been devoted to implementing low-dimensional topologically non-trivial models in clean and highly controllable systems. Topological states have been observed and characterized in light-based platforms such as photonic crystals \cite{PhotCrystal1,PhotCrystal2,PhotCrystal3,PhotCrystal4,PhotCrystal5} and photonic quantum walks \cite{QWalk1,QWalk2}. Ultracold atoms in optical lattices are also a well-suited environment to implement topological phases of matter \cite{ReviewUltracoldTop}. In one-dimensional (1D) fermionic systems, there have been proposals to dynamically probe topological edge states \cite{ProbeFermions} and to implement topological quantum walks \cite{QWalkFermion} and symmetry-protected topological phases \cite{SPT1fermion,SPT2fermion,SPT3fermion}, which have also been observed experimentally \cite{SPTexpfermion}. In 1D bosonic systems there have also been striking advances, such as the prediction of topological states in quasiperiodic lattices \cite{QuasiPeriodic1,QuasiPeriodic2}, the direct measurement of the Zak's phase \cite{ZakPhase} in a dimerized lattice \cite{MeasureZak}, the observation of the edge states \cite{SSHbosons1,SSHbosons2} of the Su-Schrieffer-Heeger (SSH) model \cite{SSHoriginal}, or the experimental realization of the topological Anderson insulator \cite{TopologicalAnderson}.
\\
\\
In this work, we consider a quasi-1D optical lattice with a diamond chain shape filled with ultracold atoms that can occupy the OAM $l=1$ local states of each site. Such a system could be experimentally realized, for instance, by exciting the atoms to the $p$-band of a conventional optical lattice \cite{HigherOrbital1,HigherOrbital2,HigherOrbital3,HigherOrbital4} or by optically transferring OAM \cite{OptOAMAtoms} to atoms confined to an arrangement of ring-shaped potentials, which can be created by a variety of techniques \cite{ring1,ring2,ring3,ring4,ring5,ring6,ring7}. At the single particle level, we show that the addition of the OAM degree of freedom makes the system acquire a topologically non-trivial nature, which is reflected in the presence of robust edge states in the energy spectrum. Similar models describing topological insulators have been introduced in \cite{similar1} and \cite{similar2}. In order to arrive at this result, we introduce and discuss in detail exact analytical mappings that allow to unravel the features of the system and to topologically characterize it. Furthermore, these mappings allow us to predict the occurrence of Aharanov-Bohm (AB) caging in the system, which consists on the confinement of specifically prepared wave packets due to quantum interference \cite{ABcageoriginal,ABcageoriginal2,ABcagephotonics,recentwork1,recentwork2}. 
\\
\\
The rest of the paper is organized as follows. In Sec.~\ref{physical} we introduce the physical system and derive the tight-binding model that we use to describe it. In Sec.~\ref{bandstructure} we compute the band structure and discuss the differences with the model of a diamond chain without the OAM degree of freedom. In Secs.~\ref{mapping1},~\ref{mapping2} and~\ref{mapping3} we introduce three successive analytical mappings that allow to understand the main features of the model such as the presence of the edge states or the Aharanov-Bohm caging effect and to fully characterize its topological nature. In Sec.~\ref{ED} we give numerical evidence of all the results derived in the previous sections by means of exact diagonalization calculations. Finally, in Sec.~\ref{conclusions} we summarize our conclusions and note some future perspectives for this work.
\section{Physical system}
\label{physical}
\begin{figure}[t!]
\centering
\includegraphics[width=\linewidth]{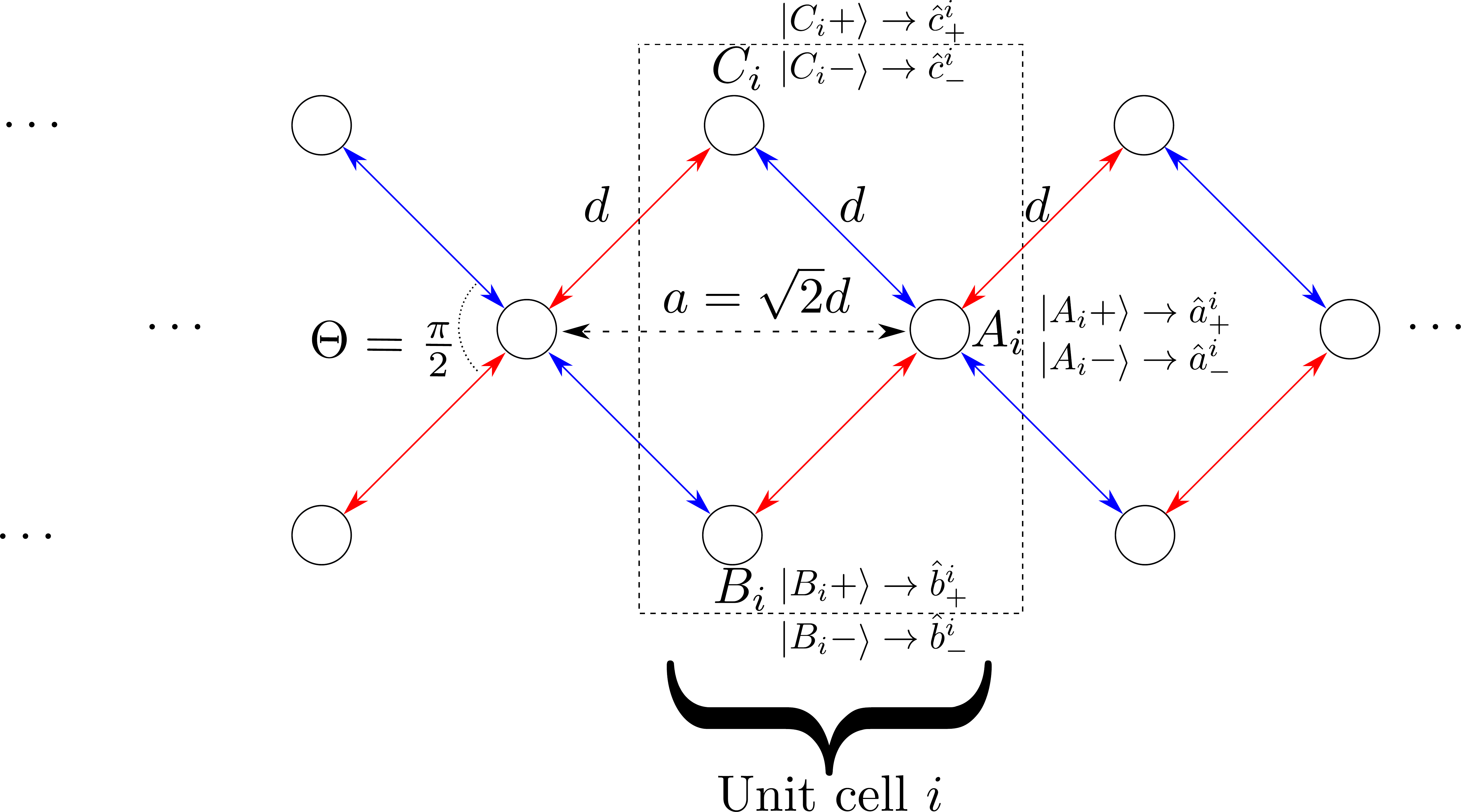}
\caption{Sketch of the diamond chain optical lattice considered in this work, indicating the shape of the unit cell as well as the two OAM $l=1$ states that can be occupied on each site and their associated operators. The directions along which all the couplings are real are signalled with blue double-arrow lines, whereas the double-arrow red lines are drawn in the directions along which the tunneling couplings involving a change in the circulation, i.e., those whose amplitude is $J_1$ or $J_3$ in \eqref{Hoperators}, acquire a $\pi$ phase.}
\label{PhySystem}
\end{figure}
\begin{figure}[t!]
\centering
\includegraphics[width=\linewidth]{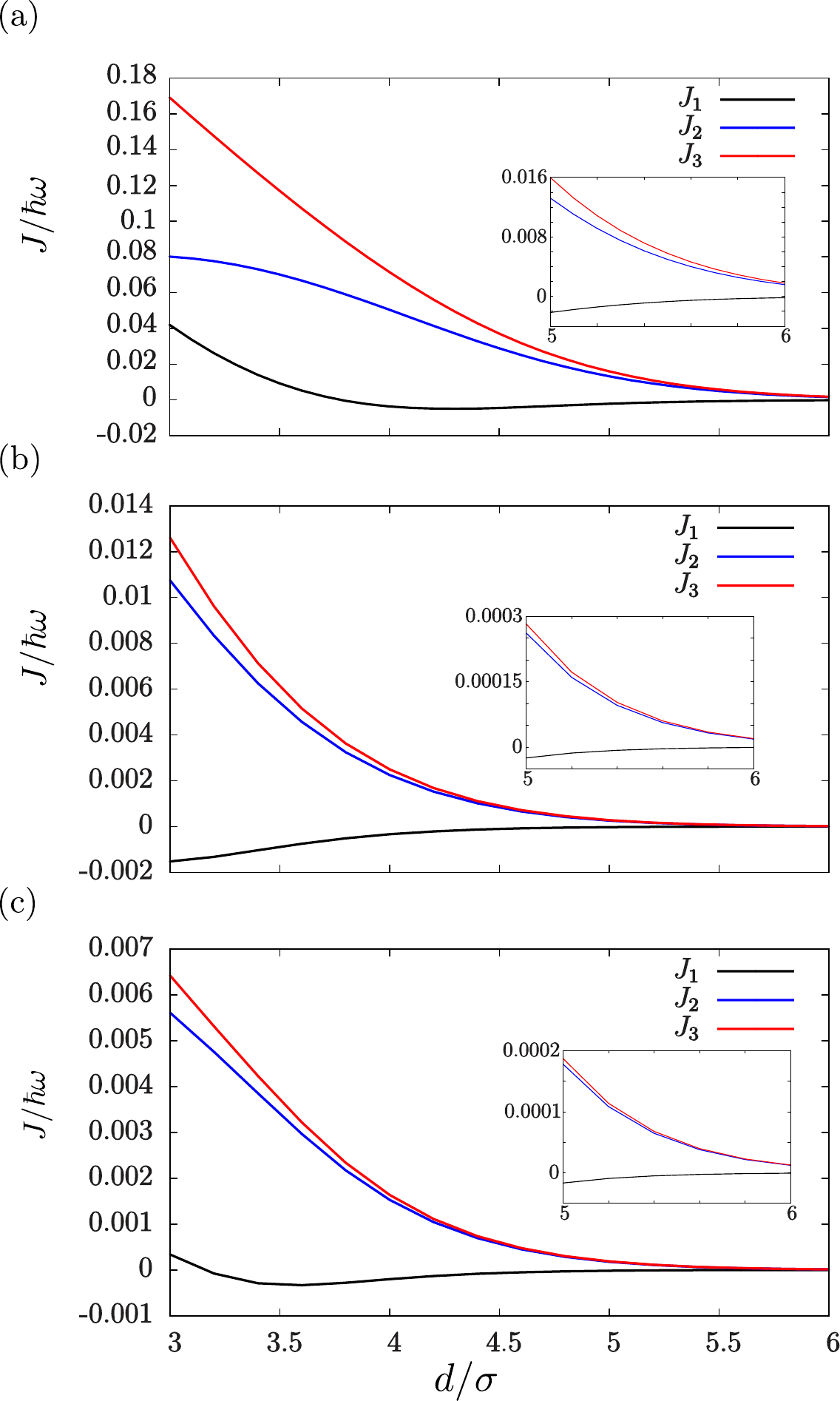}
\caption{Dependence of the $J_1$, $J_2$ and $J_3$ tunneling amplitudes on the trap separation $d$ for two ring potentials of radius (a) $R=0$ (harmonic traps), (b) $R=2.5\sigma$ and (c) $R=5.0\sigma$. In each of the plots, the insets show the values of the tunneling amplitudes for large values of $d$, where $J_2\approx J_3$ and $|J_1|\ll |J_2|,|J_3|$.}
\label{couplingsdistance}
\end{figure}
The physical system that we consider is depicted in Fig.~\ref{PhySystem}. It consists of an ultracold gas of atoms of mass $m$ trapped in a quasi-1D optical lattice with the shape of a diamond chain. The chain is formed by an integer number $N_c$ of unit cells, each of which has a central site $A$ and two sites, $B$ and $C$, equally separated from $A$ and with the lines connecting them to it forming a $\pi/2$ relative angle. Each of the sites is the center of a cylindrically symmetric potential with trapping frequency $\omega$ such as, for instance, a ring-shaped trap of radius $R$ that generates a potential $V(r)=\frac{1}{2}m\omega^2(r-R)^2$, where $r$ is the radial coordinate about the centre of the trap. In the case $R=0$, the ring trap reduces to a harmonic potential. As shown in Fig.~\ref{PhySystem}, we denote the distance between the minima of the nearest-neighbour potentials as $d$, so that the unit cells are separated by a distance $a=\sqrt{2}d$. The atoms may occupy the two states of total Orbital Angular Momentum (OAM) $l=1$ of each site, $\ket{j_i,\pm}$, where $i$ is an index labelling the unit cell and $j=A,B,C$. Thus, the total field operator of the system reads
\begin{align}
\hat{\Psi}=\sum_{i=1;}^{N_c}\sum_{\alpha=\pm}& \phi_{\alpha}^{A_i}(r_{A_i},\varphi_{A_i})\hat{a}_{\alpha}^{i}+\phi_{\alpha}^{B_i}(r_{B_i},\varphi_{B_i})\hat{b}_{\alpha}^{i}+\nonumber\\
&+\phi_{\alpha}^{C_i}(r_{C_i},\varphi_{C_i})\hat{c}_{\alpha}^{i},
\label{bosonicfield}
\end{align}
where
\begin{equation}
\phi_{\alpha}^{j_i}(r_{j_i},\varphi_{j_i})=\braket{\vec{r}|j_i,\pm}=\psi(r_{j_i})e^{\alpha i(\varphi_{j_i}-\varphi_0)}
\label{wavefunctions}
\end{equation}
are the wavefunctions of the OAM $l=1$ states with positive/negative circulation ($\alpha=+/-$) with respect to center of each site $j_i$, and $\hat{a}_{\alpha}^{i},\hat{b}_{\alpha}^{i},\hat{c}_{\alpha}^{i}$ are the annihilation operators of these states at the sites $A_i$, $B_i$ and $C_i$, respectively.
\\
We will analyze the non-interacting case, for which the Hamiltonian is 
\begin{equation}
\hat{H}=\int d\vec{r}\hat{\Psi}^{\dagger}\left[-\frac{\hbar^2\nabla^2}{2m}+V(\vec{r})\right]\hat{\Psi},
\label{HamTotal}
\end{equation}
where $V(\vec{r})$ can be taken in a good approximation as a truncated combination of all the cylindrically symmetric potentials centered at each of the sites forming the diamond chain.
\\
The Hamiltonian \eqref{HamTotal} essentially describes the tunneling dynamics of ultracold atoms between the different coupled traps of the diamond chain restricted to the manifold of $l=1$ local OAM states of each site.
This type of dynamics was studied in detail for the case of systems formed by two and three sided-coupled traps in \cite{geometricallyinduced}, where tunneling is induced by overlaps between the wave functions localized at each trap. First, by analysing the mirror symmetries of the two-trap problem, one realizes that there are only three independent tunneling amplitudes, which we will denote as $J_1$, $J_2$ and $J_3$. More specifically, $J_1$ corresponds to the self-coupling between the two OAM states of each trap induced by the breaking of the global cylindrical symmetry of the problem, $J_2$ corresponds to the cross-coupling between states in different sites with the same circulation $\alpha$, and $J_3$ corresponds to the cross-coupling between states in different sites with different circulations. As shown in Fig.~\ref{couplingsdistance}, the relative value of the three couplings depends on the inter-trap separation $d$. For short $d$, $J_3$ is appreciably larger than $J_2$, but as the distance is increased, they become closer until they take approximately the same value. Regardless of the distance, the absolute value of the self-coupling $J_1$ remains approximately one order of magnitude lower than $J_2$ and $J_3$. 
\\
In the two-trap problem, one can always take a null value for the arbitrary phase origin $\varphi_0$ of the OAM wavefunctions \eqref{wavefunctions} and thus all the tunneling couplings become real \cite{geometricallyinduced}. However, when one considers a system of three sided-coupled traps that form a triangle of central angle $\Theta$, there is a relative angle between the line defining the origin of phases and at least one of the lines connecting the centers of the traps, and thus extra phases in the tunneling amplitudes appear \cite{geometricallyinduced}. These phases are a natural consequence of the azimuthal phase present in the wavefunction of the OAM states \eqref{wavefunctions}, and can be modulated by tuning the geometry of the system, i.e. the central angle $\Theta$.
\\
Since the strength of the tunneling amplitudes decays rapidly with $d$ (see Fig.~\ref{couplingsdistance}), in the diamond chain it is a good approximation to consider coupling terms only between the nearest-neighbouring sites \cite{edgelike}. Within this approximation, if one expresses the distances and energies in harmonic oscillator (h.o.) units of $\sigma\sqrt{\hbar /(m\omega)}$ and $\hbar\omega$ respectively, and sets the origin of phases along the direction of the lines connecting the sites $C_i$, $A_i$ and $B_{i+1}$, the Hamiltonian \eqref{HamTotal} can be expressed in terms of the annihilation and creation operators as      
\begin{align}
&\hat{H}=J_1\sum_{\alpha=\pm}e^{i\pi}\hat{b}_{\alpha}^{1\dagger}\hat{b}_{-\alpha}^{1}+\hat{c}_{\alpha}^{1\dagger}
\hat{c}_{-\alpha}^{1}\nonumber\\
&+J_2\sum_{i=1}^{N_c}\sum_{\alpha=\pm}\left[\hat{a}_{\alpha}^{i\dagger}(\hat{b}_{\alpha}^{i}+\hat{b}_{\alpha}^{i+1}+\hat{c}_{\alpha}^{i}+\hat{c}_{\alpha}^{i+1})\right]+\text{h.c.}\nonumber\\
&+J_3\sum_{i=1}^{N_c}\sum_{\alpha=\pm}\left[\hat{a}_{\alpha}^{i\dagger}(e^{i\pi}\hat{b}_{-\alpha}^{i}+\hat{b}_{-\alpha}^{i+1})
+\hat{c}_{-\alpha}^{i}+e^{i\pi}\hat{c}_{-\alpha}^{i+1})\right]+\text{h.c.}
\label{Hoperators}
\end{align}
As shown in Fig.~\ref{PhySystem}, along the lines connecting the $B_i$, $A_i$ and $C_{i+1}$ sites (signalled with red double-arrow lines) the tunneling processes involving a change in the circulation acquire a $\pi$ phase. Note that the self-coupling amplitude $J_1$ is only present at the left corners of the chain. This is due to the fact that these are the only sites of the chain that are connected to only one site, whereas the rest of sites are connected to an even number of sites and, for the central angle $\theta=\pi/2$, the contributions to the self-coupling amplitude coming from the different sites interfere destructively \cite{geometricallyinduced}. We point out that the Hamiltonian \eqref{Hoperators} possesses inversion symmetry, so that the Zak's phase associated with each of the energy bands can only take the values $0$ and $\pi$ \cite{ZakPhase}.
\\
The rest of the paper will be devoted to the study and topological characterization of the spectrum and the eigenstates of the Hamiltonian \eqref{Hoperators}. Since the self-coupling at the left edge of the chain is a small effect, for simplicity we will initially take $J_1=0$ in the following sections, and then return to the case of a non-zero value for $J_1$.  
\section{Band structure}
\label{bandstructure}
\begin{figure}[t!]
\centering
\includegraphics[width=\linewidth]{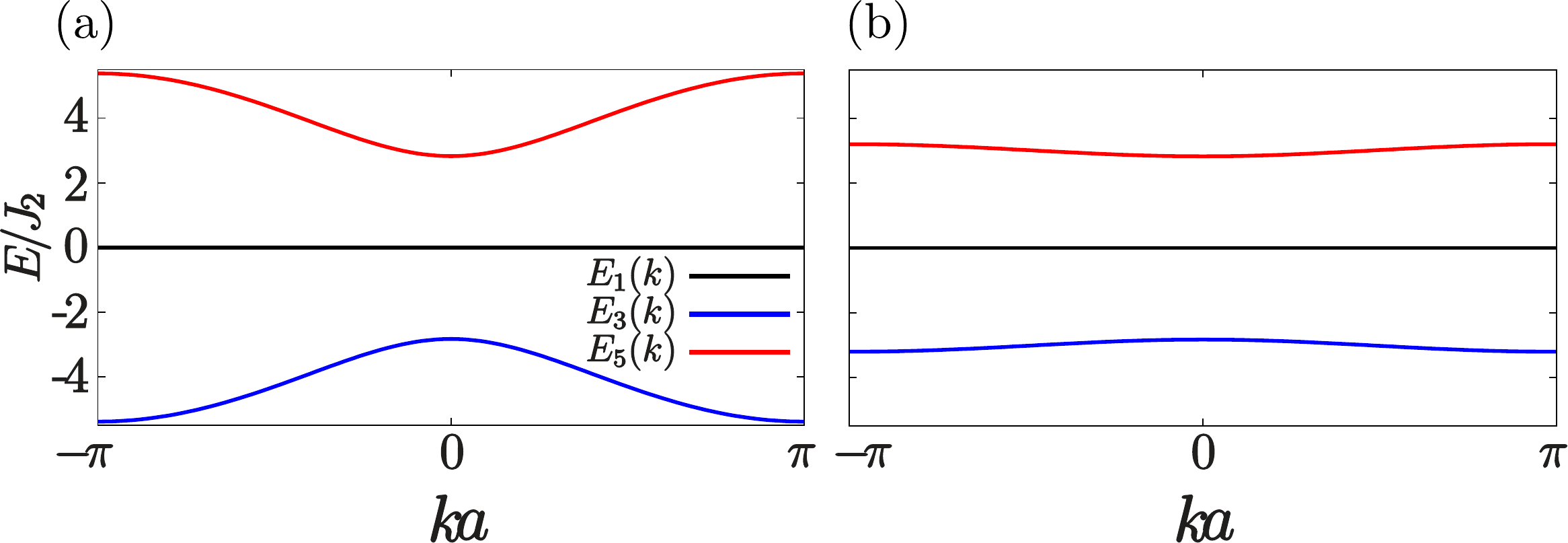}
\caption{Energy bands of the diamond chain in the OAM $l=1$ manifold computed using the values of $J_2$ and $J_3$ that are obtained for harmonic potentials separated by distances $d=3.5\sigma$ (a) and $d=6\sigma$ (b), shown in Fig.~\ref{couplingsdistance} (a).}
\label{plotbands}
\end{figure}

In order to gain a first insight into the implications of the OAM degree of freedom, we consider the limit of a large chain, $N_c\rightarrow \infty$, and compute the band structure. In order to do this calculation, we employ the usual method of taking into account the periodicity of the chain to Fourier-expand the annihilation operators as
\begin{equation}
\hat{j}_{\alpha}^i=\frac{1}{\sqrt{N_c}}\sqrt{\frac{a}{2\pi}}\int_{-\frac{\pi}{a}}^{\frac{\pi}{a}}dke^{ikx_i}\hat{j}_{\alpha}^k,
\label{FourierExpansion}
\end{equation}
where $x_i$ is the position of the $i$th cell along the direction of the diamond chain, $k$ is the quasi-momentum, $j=\{a,b,c\}$ and $\alpha=\pm$. Since there are six states per unit cell (two for each of the three sites), we obtain six energy bands. By plugging the expansion \eqref{FourierExpansion} into the Hamiltonian \eqref{Hoperators}, we can re-express it in $k$-space as 
\begin{equation}
\hat{H}^k=\int dk \Psi_k^{\dagger}H_k\Psi_k,
\end{equation}
with $\Psi_k^{\dagger}=(\hat{a}_+^k,\hat{a}_-^k,\hat{b}_+^k,\hat{b}_-^k,\hat{c}_+^k,\hat{c}_-^k)$ and
\begin{widetext}
\begin{equation}
H_k=\begin{pmatrix}
0&0&J_2(1+e^{ika})&J_3(-1+e^{ika})&J_2(1+e^{ika})&J_3(1-e^{ika})\\
0&0&J_3(-1+e^{ika})&J_2(1+e^{ika})&J_3(1-e^{ika})&J_2(1+e^{ika})\\
J_2(1+e^{-ika})&J_3(-1+e^{-ika})&0&0&0&0\\
J_3(-1+e^{-ika})&J_2(1+e^{-ika})&0&0&0&0\\
J_2(1+e^{-ika})&J_3(1-e^{-ika})&0&0&0&0\\
J_3(1-e^{-ika})&J_2(1+e^{-ika})&0&0&0&0
\end{pmatrix}.
\label{Hk}
\end{equation}
\end{widetext}
From \eqref{Hk}, it can be checked that the model possess also chiral symmetry, since the matrix ${\Gamma=\text{diag}\{-1,-1,1,1,1,1\}}$ fulfills the relation ${\Gamma H_k \Gamma=-H_k}$. The energy bands are given by the eigenvalues of the matrix $H_k$. They appear in three degenerate pairs, and are given by the expressions
\begin{subequations}
\begin{align}
E_1(k)=E_2(k)&=0\\
E_3(k)=E_4(k)&=-2\sqrt{(J_2^2+J_3^2)+\cos(ak)(J_2^2-J_3^2)}\\
E_5(k)=E_6(k)&=2\sqrt{(J_2^2+J_3^2)+\cos(ak)(J_2^2-J_3^2)}.
\end{align}
\label{energybands}
\end{subequations}
\hspace{-0.15cm}The band structure \eqref{energybands} always presents an energy gap of size $2\sqrt{2}J_2$. Two of the bands are flat regardless of the values of $J_2$ and $J_3$, and, in the $J_2=J_3$ limit, which can be realized by setting a large value of $d$, all of the six bands become flat. These facts are illustrated in Fig.~\ref{plotbands}, where we have plotted the energy bands \eqref{energybands} using realistic values of $J_2$ and $J_3$ computed for different values of $d$ and considering harmonic traps.
\\
The band structure of the diamond chain in the $l=1$ manifold presents some differences with the one that would be obtained in the manifold of ground states ($l=0$) of each of the sites. In this manifold, there is only one state per site and one tunneling amplitude $J$, which does not acquire any phases. The three energy bands that one obtains in this system are $E(k)=0,\pm 2\sqrt{2}J\cos(ka/2)$. Although there is a zero-energy band just like in the OAM $l=1$ manifold, the other two bands have always the same shape and they close at the points $k=\pm \pi/a$. However, if a flux through the plaquettes of the diamond chain is introduced, a gap opening occurs \cite{fermionsdiamondchain}. In the next section, we will perform a mapping which will more clearly demonstrate that the introduction of the OAM degree of freedom can be regarded as a net flux through the plaquettes.   
\section{Mapping into two decoupled diamond chains}
\label{mapping1}
\begin{figure}[t!]
\centering
\includegraphics[width=0.8\linewidth]{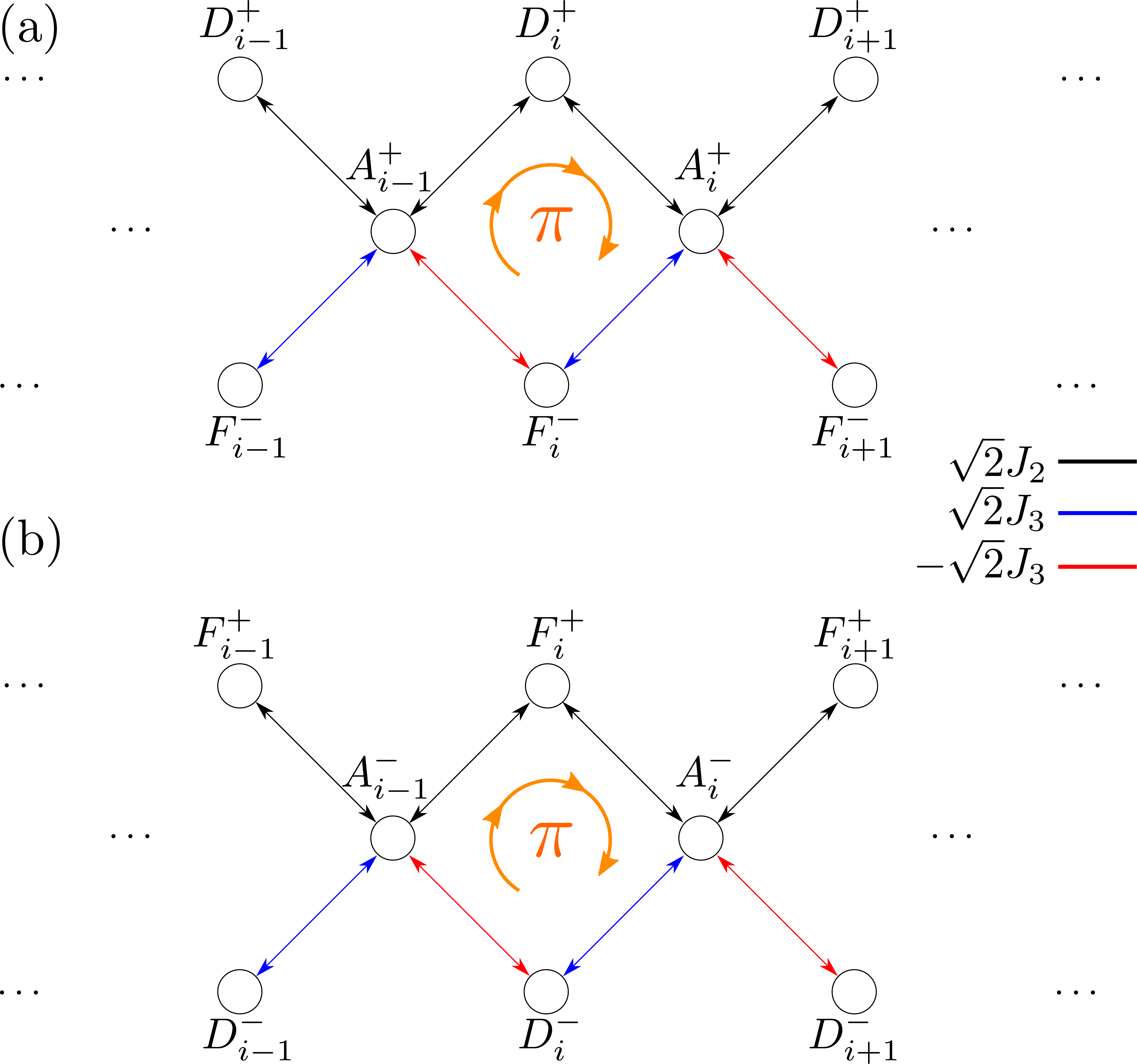}
\caption{Sketch of the two decoupled diamond chains $H^+$ (a) and $H^-$ (b) that are obtained after performing the basis rotation \eqref{1strotation}.}
\label{firstmapping}
\end{figure}

Many features of the band structure can be understood by performing exact mappings of the diamond chain in the OAM $l=1$ manifold into other models. First, let us consider the following basis rotation  
\begin{subequations}
\begin{align}
\ket{D_i,\pm}=\frac{1}{\sqrt{2}}(\ket{C_i,+}\pm\ket{B_i,+})\\
\ket{F_i,\pm}=\frac{1}{\sqrt{2}}(\ket{C_i,-}\pm\ket{B_i,-}).
\end{align}
\label{1strotation}
\end{subequations}
The only non-vanishing matrix elements of the Hamiltonian \eqref{Hoperators} in this basis are
\begin{subequations}
\begin{align}
&\braket{A_i,+|\hat{H}|D_i,+}=\braket{A_i,+|\hat{H}|D_{i+1},+}=\sqrt{2}J_2\\
&\braket{A_i,-|\hat{H}|F_i,+}=\braket{A_i,-|\hat{H}|F_{i+1},+}=\sqrt{2}J_2\\
&\braket{A_i,+|\hat{H}|F_i,-}=\braket{A_i,-|\hat{H}|D_{i},-}=\sqrt{2}J_3\\
&\braket{A_i,+|\hat{H}|F_{i+1},-}=\braket{A_i,-|\hat{H}|D_{i+1},-}=-\sqrt{2}J_3.
\end{align}
\label{couplings1strotation}
\end{subequations}
\hspace{-0.15cm}The fact that only these couplings survive after the basis rotation \eqref{1strotation} can be interpreted as a splitting of the original diamond chain with two states per site into two identical and decoupled diamond chains, one in which the $\ket{D_i,+}$ and $\ket{F_i,-}$ states are coupled to the $\ket{A_i,+}$ states and another one in which the $\ket{F_i,+}$ and $\ket{D_i,-}$ states are coupled to the $\ket{A_i,-}$ states. These two chains, labelled $H^+$ and $H^-$ respectively, are depicted in Fig.~\ref{firstmapping} and are described by the Hamiltonians
\begin{subequations}
\begin{align}
\hat{H}^{+}&=\sum_{i=1}^{N_c} \hat{a}_+^{i\dagger}[\sqrt{2}J_2(\hat{d}_+^{i}+\hat{d}_+^{i+1})+\sqrt{2}J_3(\hat{f}_-^{i}-\hat{f}_-^{i+1})]+\text{h.c.}\\
\hat{H}^{-}&=\sum_{i=1}^{N_c} \hat{a}_-^{i\dagger}[\sqrt{2}J_2(\hat{f}_+^{i}+\hat{f}_+^{i+1})+\sqrt{2}J_3(\hat{d}_-^{i}-\hat{d}_-^{i+1})]+\text{h.c.},
\end{align}
\label{HamsDecoupledfirstmapping}
\end{subequations}
\hspace{-0.15cm}where $\hat{d}_{\pm}^{i}$ and $\hat{f}_{\pm}^{i}$ are the annihilation operators associated to the states $\ket{D_i,\pm}$ and $\ket{F_i,\pm}$. Each of these two identical Hamiltonians has the same band structure \eqref{energybands} as the original one, but with three bands instead of six because there is only one state per site. This makes it possible to understand the degeneracy of the spectrum in the original model, which is a consequence of the symmetry between the two OAM states with different circulations. As shown in Fig.~\ref{firstmapping}, the fact that in each chain one of the couplings has a minus sign can be regarded as net $\pi$ flux through the plaquettes of the diamond chain. As we discussed in the previous section, this effective net flux through the plaquettes explains the gap opening in the band structure.
\section{Mapping into a modified SSH chain}
\label{mapping2}
\begin{figure}[t!]
\centering
\includegraphics[width=\linewidth]{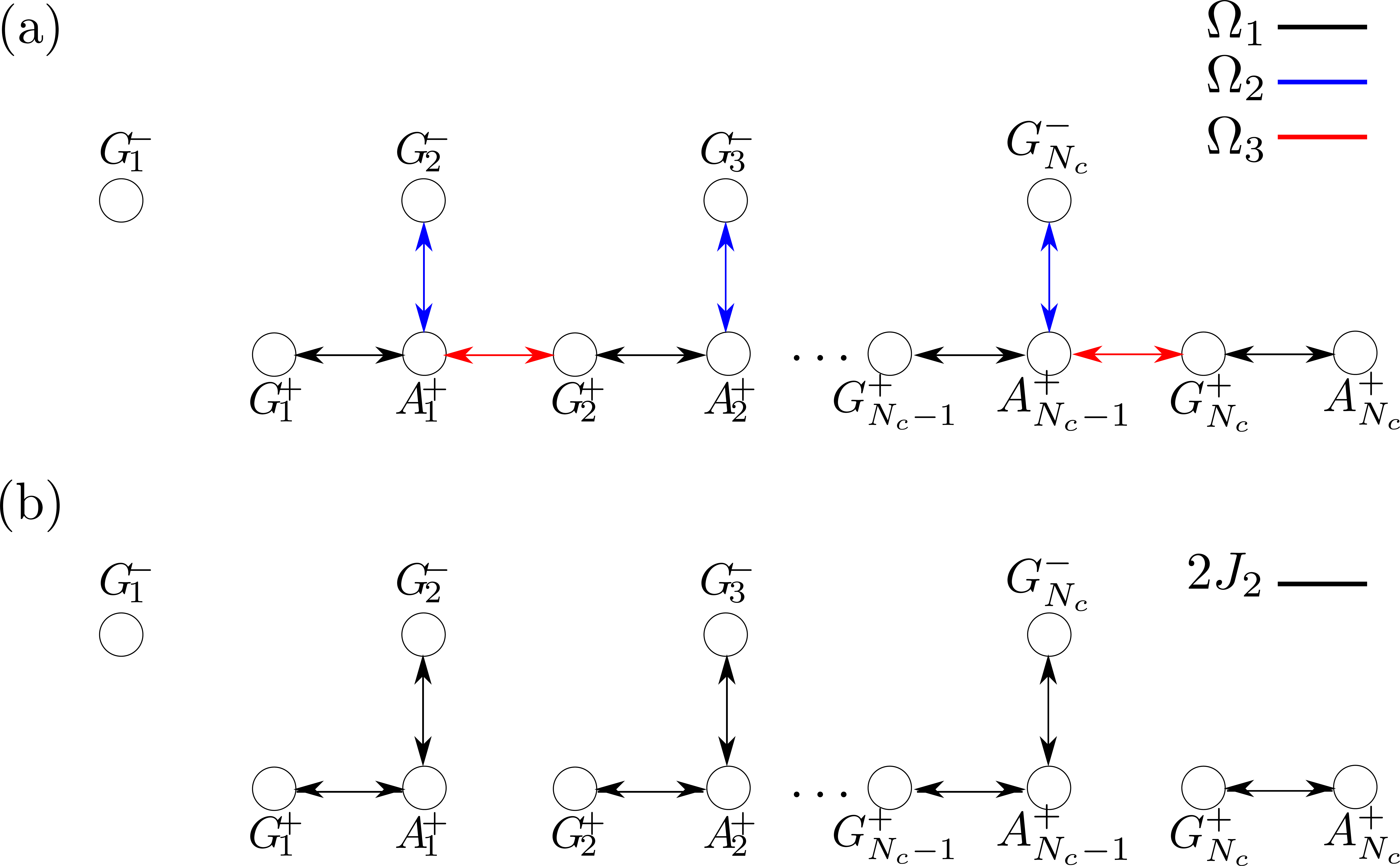}
\caption{(a) Sketch of the modified SSH chain that is obtained after performing the rotation \eqref{2ndrotation} in the $H^+$ diamond chain that was obtained after the first basis rotation \eqref{1strotation}. (b) Modified SSH chain in the $\Omega_3=0$ ($J_2=J_3$) limit, where the bulk sites become decoupled in trimers and an isolated dimer appears at the right edge.}
\label{secondmapping}
\end{figure}

We can gain further insight into the features of the band structure by performing a second basis rotation, given for the $H^+$ chain by
\begin{subequations}
\begin{align}
\ket{G_i,+}=\frac{1}{\sqrt{J_2^2+J_3^2}}(J_2\ket{D_i,+}+J_3\ket{F_i,-})\\
\ket{G_i,-}=\frac{1}{\sqrt{J_2^2+J_3^2}}(J_3\ket{D_i,+}-J_2\ket{F_i,-}).
\end{align}
\label{2ndrotation}
\end{subequations} 
For the $H^-$ chain, an equivalent mapping can be defined by substituting $F$ by $D$ everywhere in eqs. \eqref{2ndrotation}. Since the two chains are identical, from now on we will base the discussion on the $H^+$ chain and indicate the results that are obtained for the $H^-$ chain.
\\
The basis rotation \eqref{2ndrotation} reduces even further the number of non-vanishing matrix elements, which now are
\begin{subequations}
\begin{align}
&\braket{A_i,+|\hat{H}^{+}|G_i,+}=\sqrt{2}\sqrt{J_2^2+J_3^2}\equiv \Omega_1\\
&\braket{A_i,-|\hat{H}^{+}|G_{i+1},-}=\frac{2\sqrt{2}J_2J_3}{\sqrt{J_2^2+J_3^2}}\equiv \Omega_2\\
&\braket{A_i,-|\hat{H}^{+}|G_{i+1},+}=\frac{\sqrt{2}(J_2^2-J_3^2)}{\sqrt{J_2^2+J_3^2}}\equiv \Omega_3
\end{align}
\label{couplings2ndrotation}
\end{subequations}
As shown in Fig.~\ref{secondmapping} (a), the couplings \eqref{couplings2ndrotation} between the states \eqref{2ndrotation} can be represented in a graphical way as a modified Su-Schrieffer-Heeger (SSH) model, consisting of the usual SSH chain \cite{SSHoriginal} with alternating strong ($\Omega_1$) and weak ($\Omega_3$) couplings and extra dangling sites connected to the chain by $\Omega_2$.
Thus, the Hamiltonian of this modified SSH chain reads
\begin{equation}
\hat{H}^{+}_{SSH}=\sum_{i=1}^{N_c}\hat{a}_+^{i\dagger}(\Omega_1\hat{g}_+^{i}+\Omega_2\hat{g}_-^{i+1}+\Omega_3\hat{g}_+^{i+1})+\text{h.c.},
\label{HamSSH}
\end{equation}
where $\hat{g}_{\pm}^{i}$ are the annihilation operators associated to the states $\ket{G_i,\pm}$. This modified SSH model allows us to clarify the origin of the flat bands as well as that of the in-gap edge states. Next, we discuss separately these two types of states
\subsection{Flat-band states}
First, let us consider the general case $J_2\neq J_3$ and therefore $\Omega_3\neq 0$, as shown in Fig.~\ref{secondmapping} (a). In this case, two of the three energy bands of $\hat{H}^+$ are dispersive, but there is always a zero-energy flat band. This band also appears in a diamond chain in which the atoms occupy the ground state ($l=0$) manifold, so its presence is insensitive to the existence of a net flux through the plaquettes \cite{fermionsdiamondchain}. In order to better understand the flat-band states of the OAM $l=1$ manifold, let us first examine the simpler case of the ground state manifold. In that manifold, there is only one tunneling amplitude $J$ which does not acquire any phase. Hence, by imposing in each cell $i$ of the chain the condition that the site $A_i$ is not populated due to destructive interference, one finds a zero-energy eigenstate localized in $i$th unit cell, given by $\frac{1}{\sqrt{2}}(\ket{B_i}-\ket{C_i})$. In a similar fashion, in the OAM $l=1$ manifold we can find zero-energy states by imposing the destructive interference condition on the $A$ sites. In the modified SSH chain picture, this is achieved by populating appropriately in every two unit cells the states $\ket{G_i,+}$, $\ket{G_i,-}$ and $\ket{G_{i+1},-}$ in such a way that there is destructive interference and neither the $\ket{A_i,+}$ nor the $\ket{A_{i-1},+}$ states are populated. The states that fulfil this condition in every pair of consecutive unit cells are
\begin{equation}
\ket{0}_i^+=\frac{1}{\sqrt{C}}\left(\frac{\Omega_3}{\Omega_2}\ket{G_i,-}-\ket{G_i,+}+\frac{\Omega_1}{\Omega_2}\ket{G_{i+1},-}\right),
\end{equation}
where $C$ is a normalization constant. It can be readily checked that this is a zero-energy eigenstate of the Hamiltonian \eqref{HamSSH}. Additionally, at the left edge of the chain the state $\ket{G_1,-}$ is decoupled from the rest of states and therefore it is a zero-energy state too. Similarly, in the $H^-$ chain one can find a state $\ket{0}_i^-$ such that $\hat{H}^-_{SSH}\ket{0}_i^-=0$. By reverting the basis rotations \eqref{2ndrotation} and \eqref{1strotation}, one can find expressions for the states $\ket{0}_i^+$ and $\ket{0}_i^-$ in the original basis and check that they are orthogonal
\begin{subequations}
\begin{align}
\ket{0}_i^+ &=\frac{1}{\sqrt{2}\Omega_1}\left[J_3(\ket{C_{i+1},+}+\ket{B_{i+1},+}-\ket{C_{i},+}-\ket{B_{i},+})\right.\nonumber\\
&\left.+J_2(\ket{B_{i},-}+\ket{B_{i+1},-}-\ket{C_{i},-}-\ket{C_{i+1},-})\right]\\
\ket{0}_i^- &=\frac{1}{\sqrt{2}\Omega_1}\left[J_3(\ket{C_{i+1},-}+\ket{B_{i+1},-}-\ket{C_{i},-}-\ket{B_{i},-})\right.\nonumber\\
&\left.+J_2(\ket{B_{i},+}+\ket{B_{i+1},+}-\ket{C_{i},+}-\ket{C_{i+1},+})\right].
\end{align}
\label{zeroenergyoriginalbasis}
\end{subequations}
\hspace{-0.15cm}From the expressions \eqref{zeroenergyoriginalbasis}, we observe that the most compact form of the localized states quadruples in size with respect to the ground state ($l=0$) manifold (occupying 8 sites instead of the 2 in the latter case), now spanning two unit cells.
\\
As we discussed when we computed the band structure, in the $J_2=J_3$ limit (which physically corresponds to having a large inter-trap separation $d$), the two energy bands that are generally dispersive become flat with energies $E=\pm 2\sqrt{2}J_2$. The corresponding eigenstates can also be analytically derived in the modified SSH chain picture. In this particular limit, we have $\Omega_3=0$ and $\Omega_1=\Omega_2=2J_2$. Thus, as shown in Fig.~\ref{secondmapping} (b), each trio of states in two consecutive unit cells $\ket{G_i,+}$, $\ket{A_i,+}$ and $\ket{G_{i+1},-}$ becomes decoupled from the rest of the chain and forms a three-site system that can be readily diagonalized. When doing so, apart from the zero-energy state that we have already discussed one finds the two eigenstates and eigenenergies
\begin{align}
\ket{E\pm}_i^+ &=\frac{1}{2}\left(\ket{G_i,+}\pm\sqrt{2}\ket{A_i,+}+\ket{G_{i+1},-}\right);\nonumber\\ &\hat{H}^+_{SSH}(\Omega_3=0)\ket{E\pm}_i^+=\pm 2\sqrt{2}J_2\ket{E\pm}_i^+.
\label{flatbandstatesomega3zero}
\end{align}
Similarly, in the $H^-$ chain there are two states $\ket{E\pm}_i^-$ in every pair of consecutive unit cells such that ${\hat{H}^-_{SSH}(\Omega_3=0)\ket{E\pm}_i^-=\pm 2\sqrt{2}J_2\ket{E\pm}_i^-}$. By reverting again the basis rotations \eqref{1strotation} and \eqref{2ndrotation}, all these states read in the original basis
\begin{subequations}
\begin{align}
&\ket{E\pm}_i^+ =\frac{1}{4}\left[\ket{C_{i},+}+\ket{B_{i},+}+\ket{C_{i+1},+}+\ket{B_{i+1},+})
\right.\nonumber\\
&+\left.(\ket{C_{i},-}-\ket{B_{i},-}-\ket{C_{i+1},-}+\ket{B_{i+1},-}\right]\pm \frac{1}{\sqrt{2}}\ket{A_i,+}\\
&\ket{E\pm}_i^-=\frac{1}{4}\left[\ket{C_{i},-}+\ket{B_{i},-}+\ket{C_{i+1},-}+\ket{B_{i+1},-})\right.\nonumber\\
&+\left.(\ket{C_{i},+}-\ket{B_{i},+}-\ket{C_{i+1},+}+\ket{B_{i+1},+}\right]\pm\frac{1}{\sqrt{2}}\ket{A_i,-}.
\end{align}
\label{flatbandstatesomega3zero_originalbasis}
\end{subequations}
\hspace{-0.15cm}Like the zero-energy states, all these states are localized in two consecutive unit cells of the original diamond chain, but now with the difference that they do not form destructive interferences on the $A$ sites and have thus non-zero energy values.
\subsection*{Aharanov-Bohm caging}
\begin{figure}[t!]
\centering
\includegraphics[width=\linewidth]{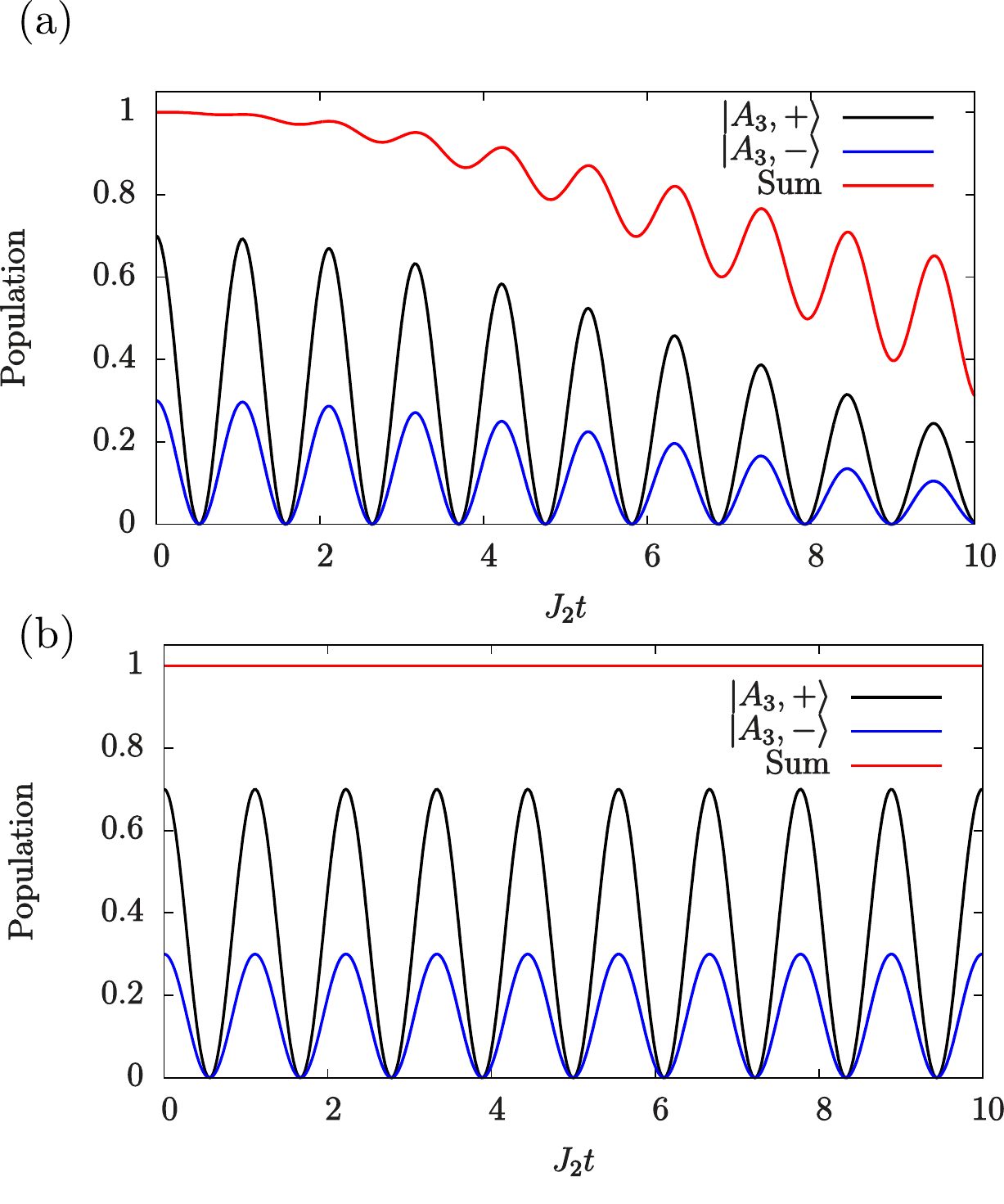}
\caption{Numerically computed time evolution, for a system of $N_c=5$ unit cells, of the population of the states $\ket{A_i,+}$ (black lines) and $\ket{A_i,-}$ (blue lines) and of the total sum of the populations of the states $\ket{B_3,\pm},\ket{C_3,\pm},\ket{A_3,\pm},\ket{B_4,\pm},\ket{C_4,\pm}$ (red lines). The tunneling parameters are (a) $J_3=1.1J_2$ and (b) $J_3=J_2$. In both cases, the initial state is $\ket{\Psi}=\sqrt{0.7}\ket{A_i,+}+\sqrt{0.3}\ket{A_i,-}$.}
\label{AharanovBohm}
\end{figure}

Aharanov-Bohm caging is a phenomenon of localization of wave packets in a periodic structure that occurs due to quantum interference. Although it was originally studied in the context of tight-binding electrons in two-dimensional lattices threaded by a magnetic flux \cite{ABcageoriginal}, its occurrence has been predicted in other physical platforms. In particular, it has been suggested and experimentally shown that Aharanov-Bohm cages can be realized in photonic lattices with a diamond-chain shape in the presence of artificial gauge fields \cite{ABcagephotonics,recentwork1,recentwork2}.
\\
In the $J_2=J_3$ limit, the system studied here also presents Aharanov-Bohm caging. In this limit, the four eigenstates \eqref{flatbandstatesomega3zero_originalbasis} are localized in the unit cells $i$ and $i+1$, forming flat bands in the spectrum of the full diamond chain. In terms of these states, the central states at site $i$, $\ket{A_i,\pm}$, can be expressed as
\begin{subequations}
\begin{align}
\ket{A_i,+}&=\frac{1}{\sqrt{2}}\left(\ket{E+}_i^+-\ket{E-}_i^+\right)\\
\ket{A_i,-}&=\frac{1}{\sqrt{2}}\left(\ket{E+}_i^--\ket{E-}_i^-\right).
\end{align}
\label{AfromFlatband}
\end{subequations}
From relations \eqref{flatbandstatesomega3zero_originalbasis} and \eqref{AfromFlatband}, we see that any initial state that is a linear combination of  $\ket{A_i,+}$ and $\ket{A_i,-}$ will evolve in time by oscillating coherently to the states $\ket{B_i,\pm}$, $\ket{C_i,\pm}$, $\ket{B_{i+1},\pm}$ and $\ket{C_{i+1},\pm}$, and therefore never populating any site beyond the unit cells $i$ and $i+1$. This Aharanov-Bohm caging effect is illustrated in Fig.~\ref{AharanovBohm}, which shows, in a system of $N_c=5$ unit cells, the numerically computed time evolution of the population of the states $\ket{A_i,+}$ (black lines) and $\ket{A_i,-}$ (blue lines) and of the total sum of the populations of the states $\ket{B_3,\pm},\ket{C_3,\pm},\ket{A_3,\pm},\ket{B_4,\pm},\ket{C_4,\pm}$ (red lines) after taking a linear combination of the $\ket{A_3,+}$ and $\ket{A_3,-}$ states. In Fig.~\ref{AharanovBohm} (a) the relation between the couplings is $J_3=1.1J_2$, so after a few oscillation periods the population escapes the cage formed by the unit cells $i$ and $i+1$. However, in the case $J_3=J_2$ plotted in Fig.~\ref{AharanovBohm} (b) the populations of the $\ket{A_3,+}$ and $\ket{A_3,-}$ states oscillate coherently without losses and the total sum of the population inside the cage remains 1 throughout the time evolution. 
\subsection{In-gap edge states in the $\Omega_3=0$ limit}
If one considers a chain of finite size, in the $\Omega_3=0$ limit discussed above there are two states at the right edge of the chain, $\ket{G_{N_c},+}$ and $\ket{A_{N_c},+}$, that are decoupled from the rest of the chain, as can be seen in Fig.~\ref{secondmapping} (b). Thus, at the right edge of the chain there are the two additional eigenstates
\begin{align}
&\ket{\text{Edge}\pm}^+=\frac{1}{\sqrt{2}}(\ket{G_{N_c},+}\pm\ket{A_{N_c},+});\nonumber\\
&\hat{H}^+_{SSH}(\Omega_3=0)\ket{\text{Edge}\pm}^+=\pm 2J_2\ket{\text{Edge}\pm}^+
\label{edgestates}
\end{align}
Similarly, in the $H^-$ chain there are edge states that fulfill $\hat{H}^-_{SSH}(\Omega_3=0)\ket{\text{Edge}\pm}^-=\pm 2J_2\ket{\text{Edge}\pm}^-$. By reverting the basis rotations \eqref{1strotation} and \eqref{2ndrotation}, we find the following expressions for all these states in the original basis
\begin{subequations}
\begin{align}
\ket{\text{Edge}\pm}^+&=\frac{1}{2\sqrt{2}}\left(\ket{C_{N_c},+}+\ket{B_{N_c},+}+\ket{C_{N_c},-}\right.\nonumber\\
&-\left.\ket{B_{N_c},-}\pm 2\ket{A_{N_c},+}\right)\\
\ket{\text{Edge}\pm}^-&=\frac{1}{2\sqrt{2}}\left(\ket{C_{N_c},-}+\ket{B_{N_c},-}+\ket{C_{N_c},+}\right.\nonumber\\
&-\left.\ket{B_{N_c},+}\pm 2\ket{A_{N_c},-}\right)
\end{align}
\end{subequations}
 Since the energies of the flat band states are $\pm 2\sqrt{2}J_2$, these edge states appear as in-gap states in the energy spectrum, which is suggestive of a possible topological origin. In order to see if the model is indeed topologically non-trivial, we should compute the Zak's phases of the different bands \cite{ZakPhase}. However, this is not possible in the original model \eqref{Hoperators} due to the degeneracy of the bands \eqref{energybands}. In the mapped models \eqref{HamsDecoupledfirstmapping} and \eqref{HamSSH} the bands are no longer degenerate, but there is no inversion symmetry and thus the Zak's phase is not quantized. It is therefore necessary to perform a third mapping into an inversion-symmetric model in order to recover quantized Zak's phases for the bands, therefore allowing for a topological characterization of the model.  
\section{Third mapping into a modified diamond chain and topological characterization}
\label{mapping3}
\begin{figure}[t!]
\centering
\includegraphics[width=\linewidth]{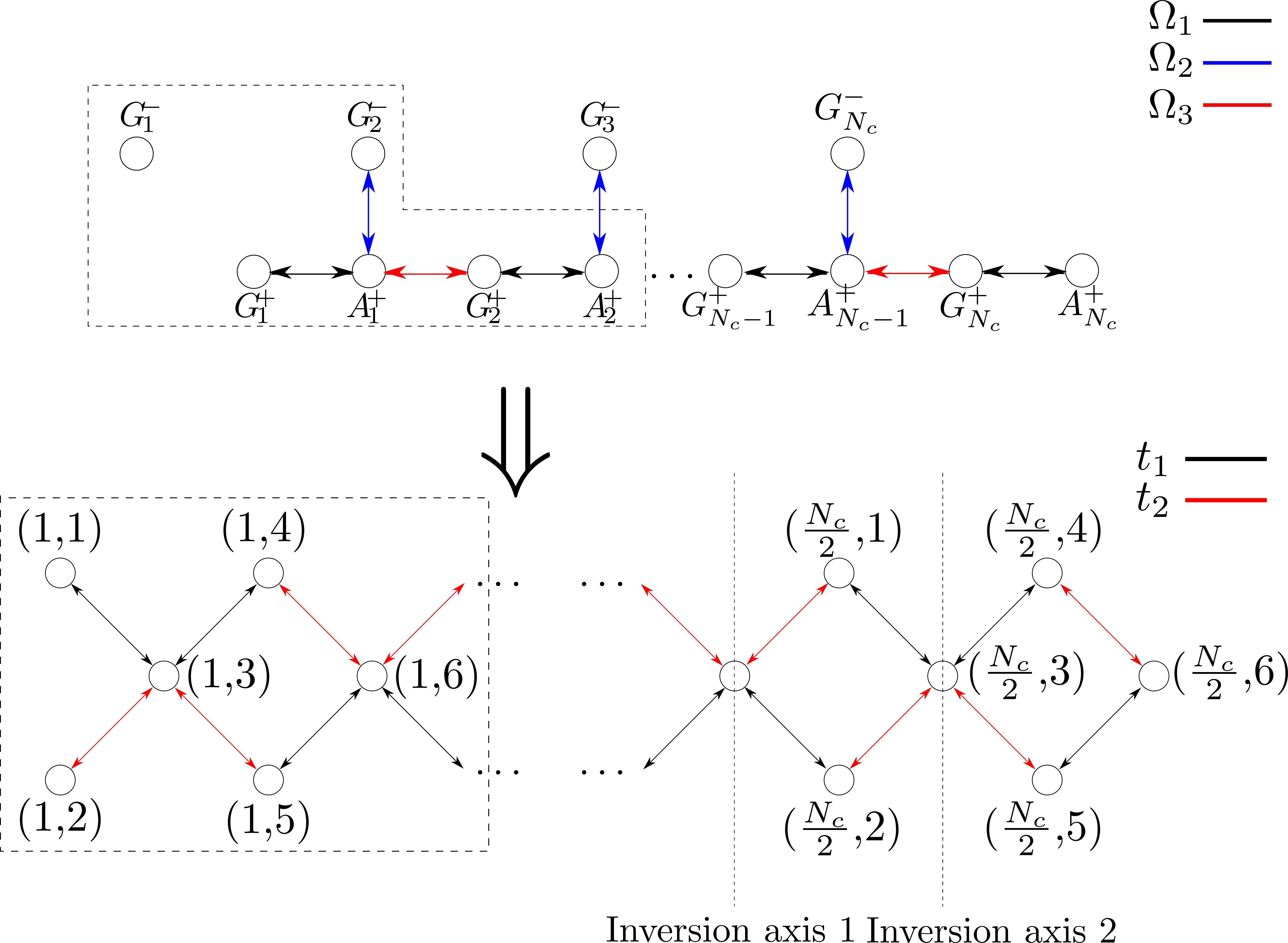}
\caption{Schematic representation of the mapping from the modified SSH $H^+$ chain into a diamond chain with alternate hoppings. The two possible choices for the inversion symmetry axis within a given unit cell, under periodic boundary conditions, are also shown. Note that neither of them are localized at the central axis of the unit cell.}
\label{thirdmapping}
\end{figure}

In order to map the modified SSH chain into a model that allows to compute meaningful Zak's phases, we take two consecutive unit cells, $i$ and $i+1$, of this relabelled chain and define a basis rotation into 6 new states, which we shall denote as $\ket{i,j}$ ($j=1,...,6$), in the following way
\begin{subequations}  
\begin{align}
&\ket{i,1}=\frac{1}{\sqrt{t_1^2+t_2^2}}(t_2\ket{G_{i+1},+}-t_1\ket{G_{i+1},-})\\
&\ket{i,2}=\frac{1}{\sqrt{t_1^2+t_2^2}}(t_1\ket{G_{i+1},+}+t_2\ket{G_{i+1},-})\\
&\ket{i,3}=\ket{A_{i+1},+}\\
&\ket{i,4}=\frac{1}{\sqrt{t_1^2+t_2^2}}(t_2\ket{G_i,+}-t_1\ket{G_i,-})\\
&\ket{i,5}=\frac{1}{\sqrt{t_1^2+t_2^2}}(t_1\ket{G_i,+}+t_2\ket{G_i,-})\\
&\ket{i,6}=\ket{A_i,+},
\end{align}
\label{3rdrotation}
\end{subequations}
where the parameters $t_1$ and $t_2$ fulfill the relations $2t_1t_2=\Omega_1\Omega_2$ and $t_1^2-t_2^2=\Omega_1\Omega_3$. After applying this rotation, a modified SSH chain of $N_c$ unit cells gets mapped into a modified diamond chain of $N_c/2$ unit cells with 6 states per unit cell and alternate $t_1$ and $t_2$ hopping constants. The resulting chain has an integer or half-integer number of unit cells depending on the parity of $N_c$. However, since there is no qualitative difference between the two cases, from now on we restrict ourselves to the case when $N_c$ is even. The mapping process and the resulting modified diamond chain are illustrated in Fig.~\ref{thirdmapping}. Note also that, under this mapping, the number of bands gets doubled (6 instead of 3) but the Brillouin zone is folded in half, such that one has the same number of allowed energy states before and after the mapping, as expected.
The Hamiltonian describing this modified diamond chain reads
\begin{align}
\hat{H}^{t_1t_2}&=\sum_{i=1}^{N_c/2}\hat{a}_i^{2\dagger}(t_1\hat{a}_{i}^{1}+t_2\hat{a}_{i}^{4})+\hat{a}_i^{3\dagger}(t_2\hat{a}_{i}^{1}+t_4\hat{a}_{i}^{4})+\text{h.c.}\nonumber\\
&+\sum_{i=1}^{N_c/2}\hat{a}_i^{5\dagger}(t_2\hat{a}_{i}^{4}+t_1\hat{a}_{i+1}^{1})+\hat{a}_i^{6\dagger}(t_1\hat{a}_{i}^{4}+t_2\hat{a}_{i+1}^{1})+\text{h.c.},
\label{HamModifiedDiamond}
\end{align} 
where $\hat{a}_i^j$ are the annihilation operators associated to the states $\ket{i,j}$ (j=1,...,6). The modified diamond chain \eqref{HamModifiedDiamond} has inversion symmetry, and thus the Zak's phases of its band are quantized. The topological characterization of this model was addressed in \cite{topologicalSSHring}. As shown in Fig.~\ref{thirdmapping}, the inversion symmetry axes are not in the center of the unit cells, and it is thus necessary to use a generalized formula to compute the Zak's phases of the bands \cite{ZakGeneralization}. Taking this issue into account, in \cite{topologicalSSHring} it was shown that the model of the modified diamond chain with alternate hoppings hosts topologically protected edge states. Thus, by reverting the mapping we can conclude that the edge states of the original diamond chain in the OAM $l=1$ manifold \eqref{Hoperators} are topologically protected, so we expect them to be robust against changes in the condition $J_2=J_3$, and to disappear completely only on the gap closing points $J_2/J_3=0$.
\section{Exact diagonalization results}
\label{ED}
\begin{figure}[t!]
\centering
\includegraphics[width=\linewidth]{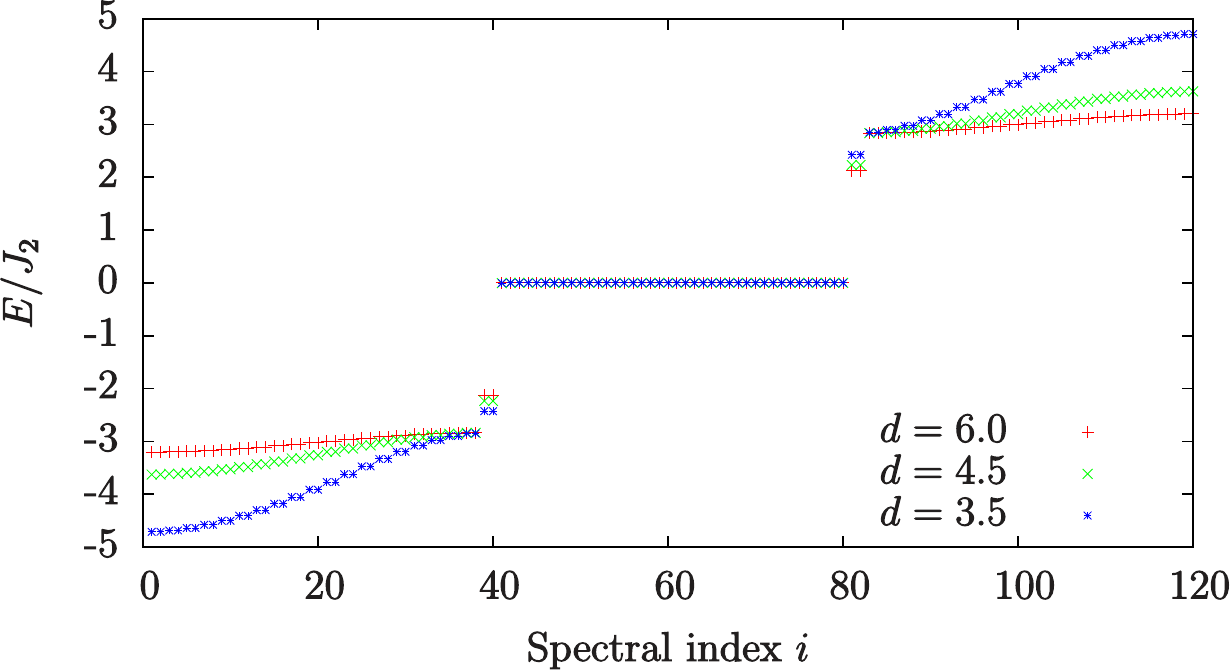}
\caption{Energy spectra of the single-particle diamond chain Hamiltonian \eqref{Hoperators} obtained with the values of $J_2$ and $J_3$ corresponding to harmonic traps separated by distances $d=$ 3.5, 4.5, and 6.0 (in h.o. units). The number of unit cells considered is $N_c=20$.}
\label{figure_spectra}
\end{figure}

\begin{figure}[t!]
\centering
\includegraphics[width=\linewidth]{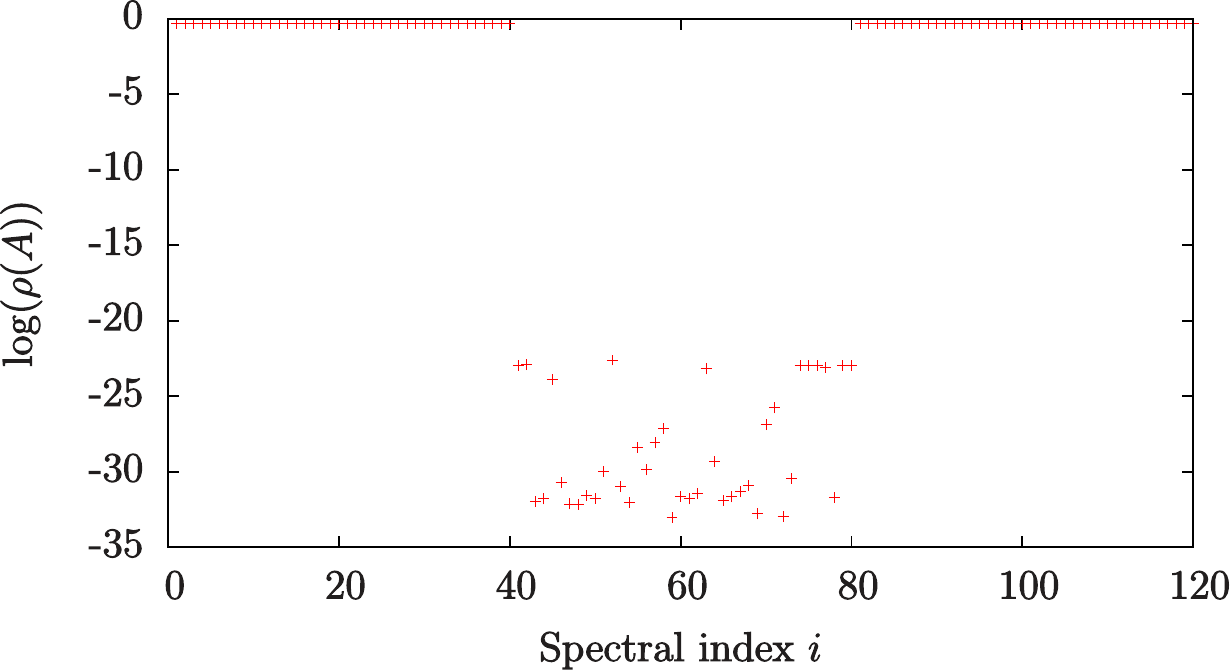}
\caption{Sum of the populations of all the states localized on the $A$ sites for the different eigenstates of the single-particle diamond chain Hamiltonian \eqref{Hoperators}, for $J_2=J_3$ and $N_c=20$ unit cells.}
\label{centralpops_60sites}
\end{figure}

\begin{figure}[t!]
\centering
\includegraphics[width=\linewidth]{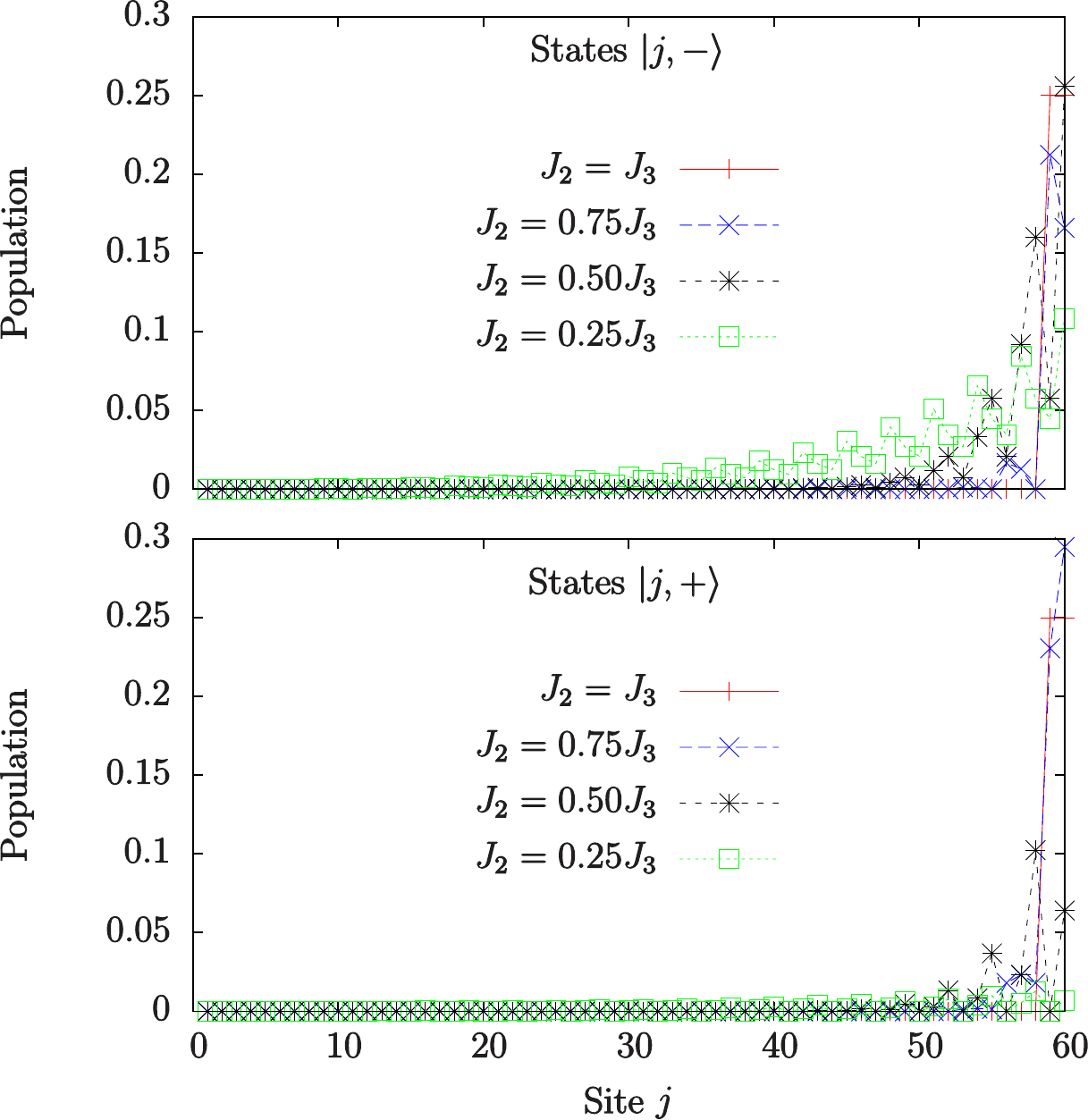}
\caption{Density profiles of one of the in-gap states (spectral index $i=39$) of a diamond chain with a total number of unit cells $N_c=20$ computed for different relative values of $J_2$ and $J_3$. The upper plot corresponds to the density distribution of the states with negative circulation and the lower plot to the states of positive circulation.}
\label{edgestate39_60sites}
\end{figure}

In order to numerically confirm the predictions that we have done in the previous sections through the band structure calculations and a series of exact analytical mappings, we have performed numerical diagonalization of the original single-particle Hamiltonian of the diamond chain \eqref{Hoperators} to find its energy spectrum and the corresponding eigenstates. We have considered a chain with $N_c=20$ unit cells ($N=3N_c=60$ sites), which has a Hilbert space of dimension $\dim{\mathcal{H}}=2N=120$, where the factor of 2 comes from the internal OAM degree of freedom. All relevant features of the model are captured by a system of this size.
\\
In Fig.~\ref{figure_spectra} we show the energy spectra that one obtains by considering the values of $J_2$ and $J_3$ corresponding to realistic calculations done with harmonic traps separated by different distances $d$. We observe that independently of the value of $d$, all energies appear in degenerate pairs as a consequence of the symmetry between the OAM $l=1$ states with different circulations. As predicted by the band structure calculation \eqref{energybands}, for all the relative values of $J_2$ and $J_3$ there is a set of states with zero energy. In all cases, we observe that these states have no population in the central sites. This fact is illustrated in Fig.~\ref{centralpops_60sites}, where we have plotted the logarithm of the total population of the states at the $A$ sites $\rho(A)$, observing a dramatic drop for the states belonging to the flat region of the spectrum.
\\
As can be seen in the inset of Fig.~\ref{couplingsdistance} (a), as one increases the distance between the traps $d$, the values of $J_2$ and $J_3$ converge, leading to a progressive flattening of the dispersive bands, as predicted by the expressions of the energy bands \eqref{energybands}. In Fig.~\ref{figure_spectra} we observe that, even though the $J_2=J_3$ limit is never reached, there always appear 4 in-gap states, which have a correspondence with the two edge states present for each of the $H^+$ and $H^-$ modified SSH chains \eqref{HamSSH}. For the case $d=6.0$, which is very close to the $J_2=J_3$ limit, these in-gap states have almost the energies $\pm 2J_2$ that we predicted when analysing the SSH chain, and as the relative difference between $J_2$ and $J_3$ is increased (i.e., as $d$ is decreased), the absolute value of the energies of these states increases. 
\\
In order to check the prediction that, due to topological protection, these states should be localized at the right edge of the chain even when $J_2$ and $J_3$ differ, we have computed their density profiles for different relative values of $J_2$ and $J_3$. The results are shown in Fig.~\ref{edgestate39_60sites}. The sites have been assigned a number $j$ according to the correspondences $C_i=3i-2$, $B_i=3i-1$, $A_i=3i$, i.e., the site $j=1$ is the $C$ site of the cell $i=1$ and the site $j=60$ is the $A$ site of the cell $i=20$. We observe that in all cases the population of both the states with negative and positive circulation is exponentially localized at the right edge of the chain. As expected, as the ratio $J_2/J_3$ deviates from 1 the edge states grow longer tails into the bulk. However, as can be seen in Fig.~\ref{edgestate39_60sites}, even in the case $J_2=0.25J_3$ the distribution shows a sharp decay into the bulk. In realistic implementations, the case that deviates most from $J_2=J_3$ would correspond to very close harmonic traps, as can be seen in Fig.~\ref{couplingsdistance} (a). But even in that case, one would have an approximate relation between the couplings $J_2\approx 0.5 J_3$, so one would observe narrowly localized edge states.      

\subsection*{Effect of $J_1$ at the edges}
\begin{figure}[t!]
\centering
\includegraphics[width=\linewidth]{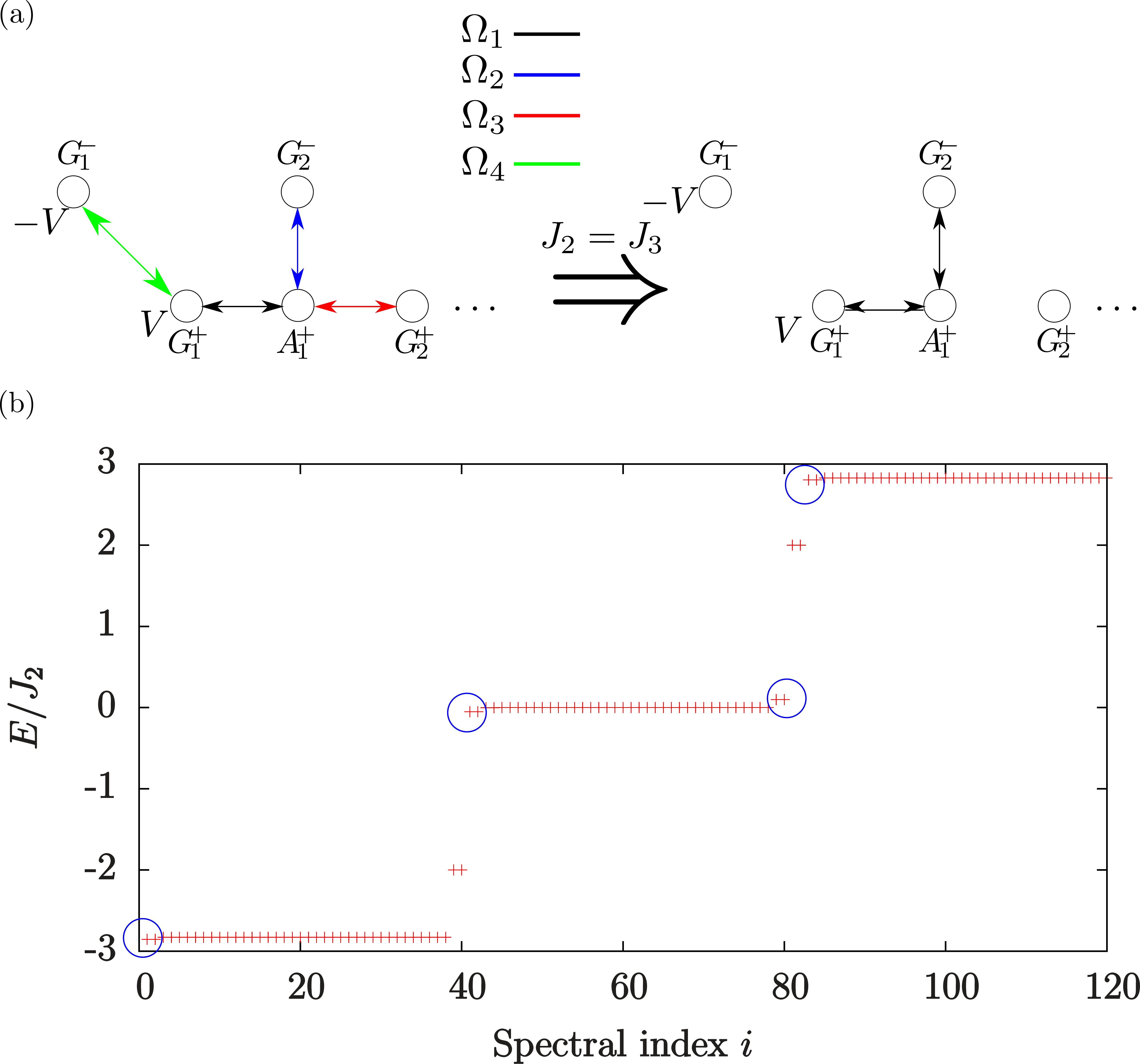}
\caption{(a) Schematic representation of the left end of the modified SSH chain in the presence of a non-zero value of $J_1$ for $J_2\neq J_3$ (left) and $J_1$ for $J_2=J_3$ (right). In the latter case, the on-site potential $V$ shifts the energies of 4 states with respect to the case $J_1=0$. (b) Energy spectrum of a diamond lattice with 20 unit cells and tunneling parameters $J_2=J_3=-10J_1$. A total of 8 states, which are indicated on the plot by blue circles, have small shifts with respect to the case $J_1=0$.}
\label{J1noZero}
\end{figure}

So far, we have neglected the effect of the self-coupling $J_1$ at the left edge of the chain, since typically $|J_1|\ll |J_2|,|J_3|$ and the self-coupling term is only present at two sites. However, this term can be readily incorporated in the exact diagonalization scheme and its effect characterized.
\\
Before presenting the numerical result, let us retrieve the effect of the self-coupling term on the analytical mappings \eqref{1strotation} and \eqref{2ndrotation}. Due to this term, the left edges of the $H^+$ and $H^-$ chains are coupled because of the matrix elements $\braket{D_1,+|\hat{H}|F_1,-}=\braket{D_1,-|\hat{H}|F_1,+}=J_1$. In the modified SSH chain obtained with the basis rotation \eqref{2ndrotation}, this extra term translates into a coupling at the left end of the chain,
\begin{equation}
\Omega_4=\braket{G_1,-|\hat{H}|G_1,+}=\frac{J_1(J_2^2-J_3^2)}{J_2^2+J_3^2}
\end{equation}  
and an on-site potential also in the two sites at the left end of the chain
\begin{equation}
V=\braket{G_1,+|\hat{H}|G_1,+}=-\braket{G_1,-|\hat{H}|G_1,-}=\frac{2J_1J_2J_3}{J_2^2+J_3^2}\approx J_1
\end{equation}
These two extra terms are illustrated in Fig.~\ref{J1noZero} (a). In the $J_2=J_3$ limit, the state $\ket{G_1,-}$ is an eigenstate of energy $-V$, and, if $|J_1|\ll |J_2|,|J_3|$, due to the on-site potential the energies of the isolated three-state system ${\ket{G_1,+},\ket{A_1,+},\ket{G_2,-}}$ are approximately $\pm 2\sqrt{2}J_2\mp V/4$ and $V/2$. In Fig.~\ref{J1noZero} (b) we show the spectrum corresponding to the tunneling amplitudes $J_2=J_3=-10J_1$, with $J_1,V<0$. Due to the contributions from the $H^+$ and $H^-$ chains, we observe that two states have energy $-V$, another two have energy $V/2$ and two states from each of the flat bands are shifted by a quantity $\approx V/4$. In summary, since the self-coupling is only present in two of the sites of the chain and its amplitude is typically much lower than the one of the cross-couplings, its effect is only to shift a few states by a small quantity and can thus be safely neglected in a diamond chain with a large number of unit cells. 
\section{Conclusions}
\label{conclusions}
In this work, we have explored the consequences of the addition of the OAM degree of freedom in the physics of ultracold atoms in optical lattices with a tunable geometry. Specifically, we have analysed the case of a single atom in a diamond-chain optical lattice, which is a simple geometry in which, due to the misalignment between the lines connecting the different sites, $\pi$ phases appear naturally in some tunneling amplitudes between the states of the OAM $l=1$ manifold. The appearance of these phases has deep consequences in the physics of the diamond chain. When one considers the case where the atom can occupy only the ground state of each trap, the band structure is gapless. By means of band structure calculations and a series of exact mappings, we have shown that when adding the OAM degree of freedom a gap opens in the spectrum and topologically protected edge states appear. We have also performed exact diagonalization calculations that support and confirm all the analytically derived results.
\\
Possible extensions of this work include considering the effect of interactions in a scenario with few or many interacting atoms or exploring the consequences and possible topological implications of the geometrically induced tunneling amplitudes for ultracold atoms carrying OAM in lattices of different geometries, and also investigating more general out-of-equilibrium dynamics in these topological systems.
\acknowledgements
{GP, JM and VA gratefully acknowledge financial support
through  the  Spanish  Ministry  of  Science  and  Innovation (MINECO) (Contract No. FIS2014-57460P) and the Catalan Government (Contract No.  SGR2014-1639). GP acknowledges inancial support from MINECO through Grant No. BES-2015-073772. AMM acknowledges financial support from the Portuguese Institute for Nanostructures, Nanomodelling and Nanofabrication (i3N) through the grant BI/UI96/6376/2018. Work at the University of Strathclyde was supported by the EPSRC Programme Grant
DesOEQ (EP/P009565/1). We thank Alessio Celi, Alexandre Dauphin, Anton Buyskikh and Stuart Flannigan for helpful discussions.

\end{document}